# RECENT MARS15 DEVELOPMENTS: NUCLIDE INVENTORY, DPA AND GAS PRODUCTION*†

N.V. Mokhov#, Fermilab, Batavia, IL 60510, U.S.A.

## Abstract

Recent developments in the MARS15 code are described for the critical modules related to demands of hadron and lepton colliders and Megawatt proton and heavy-ion beam facilities. Details of advanced models for particle production and nuclide distributions in nuclear interactions at low and medium energies, energy loss, atomic displacements and gas production are presented along with benchmarking against data.

*Work supported by Fermi Research Alliance, LLC under contract No. DE-AC02-07CH11359 with the U.S. Department of Energy.
†Presented paper at 46th ICFA Advanced Beam Dynamics Workshop on High-Intensity and High-Brightness Hadron Beams, Sept. 27 - Oct. 1, 2010, Morschach, Switzerland.
#mokhov@fnal.gov

# RECENT MARS15 DEVELOPMENTS: NUCLIDE INVENTORY, DPA AND GAS PRODUCTION*

N.V. Mokhov#, Fermilab, Batavia, IL 60510, U.S.A

*Abstract*

Recent developments in the MARS15 code are described for the critical modules related to demands of hadron and lepton colliders and Megawatt proton and heavy-ion beam facilities. Details of advanced models for particle production and nuclide distributions in nuclear interactions at low and medium energies, energy loss, atomic displacements and gas production are presented along with benchmarking against data.

## PHYSICS MODEL DEVELOPMENTS

The focus of recent developments to the MARS15 code [1] was on particle production in nuclear interactions at low and medium energies crucial for an accurate description of radiation effects in numerous applications at particle colliders and high-power beam facilities [2]. Substantial improvements have been done to the MARS code event generator LAQGSM [3], the quark-gluon string model. These include low-energy projectiles (p, γ, and heavy ions), near-threshold kaon production, low-energy pions for precision experiments and a neutrino factory, inverse reactions, cross-sections of light nuclear projectiles, light target nuclei (hydrogen, deuterium, and tritium), machine-independent form, and thorough tests on various platforms. An example of benchmarking [4] is shown in Fig. 1 for a quite difficult case of kaon production in the near-threshold region on deuterium in comparison to the ANKE spectrometer data at COSY-Julich [5]. The agreement is amazingly good.

Electromagnetic shower (EMS) module – crucial in energy deposition calculations practically in all applications – has been substantially extended. The exclusive and hybrid modelling options have been added for all EMS processes and all photo- and electro-nuclear hadron and muon production reactions with user-controlled material-dependent switches between the exclusive, inclusive and hybrid modes. The appropriate choice of these parameters substantially reduces variance and improves a computational efficiency.

Electromagnetic interactions of heavy-ion beams, recoil nuclei and fragments generated in nuclear interactions are, in many cases, the most important contributor to radiation effects. In MARS15, knock-on electron production above the material/projectile-dependent thresholds is accurately modelled, with remaining (restricted) energy loss treated continuously down to 1 keV. The ionization energy loss model for an arbitrary projectile has been further updated [6] with a modified Thomas-Fermi expression for ion effective charge based on that by Pierce and Blann, and taking into account available information on probabilities of different ion charge states for few-electron heavy ions at intermediate energies. Fig. 2 shows calculated dE/dx *vs* data for various particles in silicon at energy 1 keV/A to 1 GeV/A.

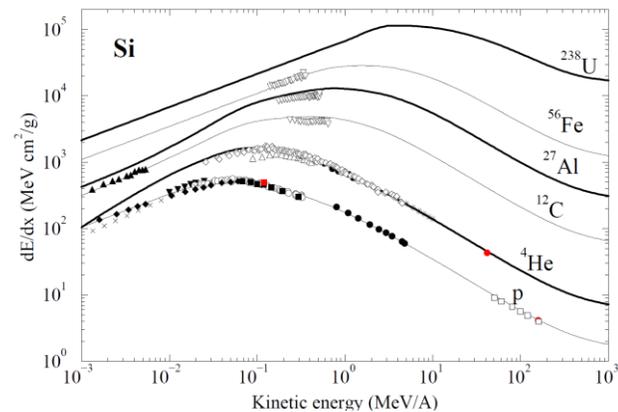

Figure 2: Ionization energy loss of projectiles from protons to uranium in silicon.

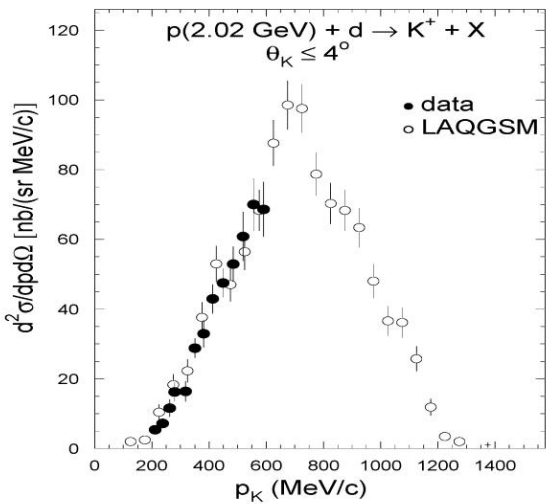

Figure 1: Double differential cross sections of produced $K^+$ mesons in interactions of protons with deuterium.

Other developments to MARS15 include a new flux-to-dose conversion module based on recent publications [7-10]; user-defined irradiation and cooling times for residual dose; extended Graphical-User Interface (GUI); extended geometry module for a higher accuracy in very complex configurations in a presence of arbitrary magnetic and electric (RF) fields; extended lists of built-in materials and nuclide distributions; adjustments to run the code in a machine-independent fashion on 32- and 64-bit platforms.

___________________________________________
*Work supported by Fermi Research Alliance, LLC under contract No. DE-AC02-07CH11359 with the U.S. Department of Energy.
#mokhov@fnal.gov

## NUCLIDE INVENTORY

Nuclide treatment in MARS15 is done in two steps: generation of nuclide A/Z distributions in the specified materials/regions followed by decay and transmutation of generated nuclides. Modelling of nuclide production is done automatically at each inelastic nuclear interaction vertex. The code event generator, based in this part on CEM and LAQGSM models [3], provides a reliable description here. An example of benchmarking of the mass yield in $^{86}$Kr + $^{9}$Be interactions at 1 GeV/A is shown in Fig. 3. Hydrogen and helium gas production tables and histograms are also a part of the MARS15 standard output.

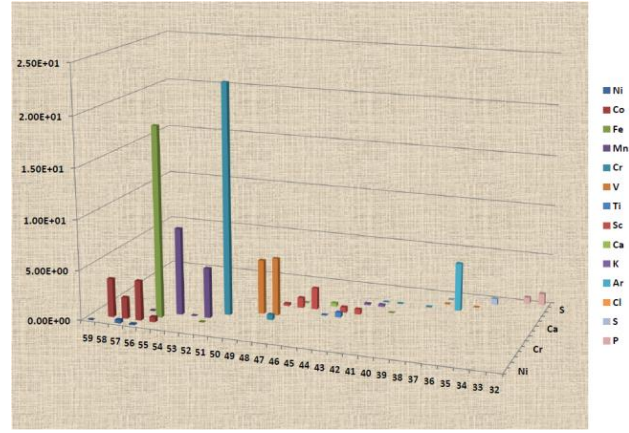

Figure 4: Calculated activity (a.u.) of the Main Injector steel collimator. For light nuclides, the activity is less than 2 a.u.

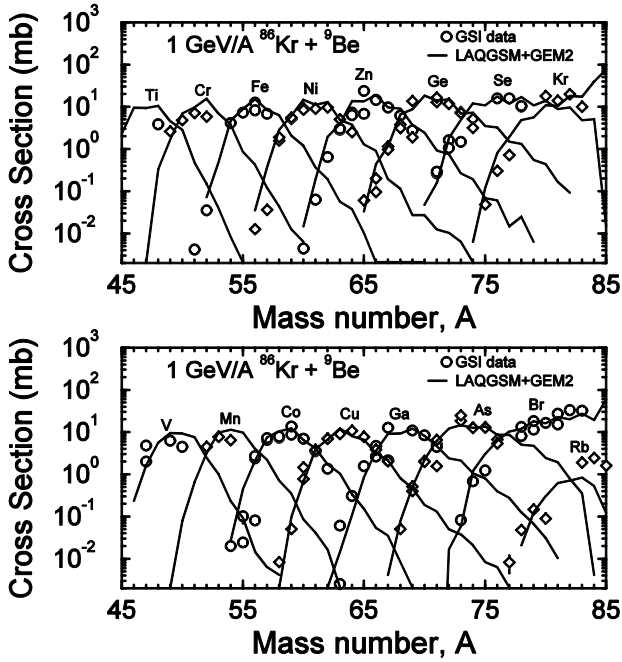

Figure 3: Calculated mass yield in $^{86}$Kr + $^{9}$Be reaction at 1 GeV/A in comparison with GSI data [11].

The second step is done using the DeTra code [12]. DeTra solves analytically the Bateman equations governing the decay, build-up and transmutation of radionuclides. The complexity of the chains and the number of nuclides are not limited. The nuclide production terms considered include transmutation of the nuclides inside the chain, external production, and fission. Time-dependent calculations are possible since all of the production terms can be re-defined for each irradiation step. The number of irradiation steps and output times is unlimited. DeTra is thus able to solve any decay and transmutation problem as long as the nuclear data, i.e. decay data and production rates, or cross sections, are known.

An example of nuclide inventory is shown in Fig. 4. It was generated for the Fermilab Main Injector collimator after 1-year irradiation by an 8-GeV proton beam and 1-day cooling.

## DISPLACEMENTS PER ATOM

Radiation damage is the displacement of atoms from their equilibrium position in a crystalline lattice due to irradiation with formation of interstitial atoms and vacancies in the lattice. Resulting deterioration of material (critical) properties is measured – in the most universal way – as a function of displacements per target atom (DPA). DPA is a strong function of projectile type, energy and charge as well as material properties including its temperature. The phenomenon becomes very serious for high-intensity beams especially for high-charge heavy ions ($\sim z^2$), being identified, for example at FRIB and FAIR, as one of the critical issues, limiting the lifetime of targets to as low as a few weeks.

DPA modelling in the MARS15 code is done for an arbitrary projectile of energy ranging from 1 keV to 10 TeV as described in Ref. [13]. A primary knock-on atom (PKA) created in nuclear collisions can generate a cascade of atomic displacements. This is taken into account via a damage function $\nu(T)$. DPA is expressed in terms of a damage cross section $\sigma_d$:

$$\sigma_d(E) = \int_{T_d}^{T_{max}} \frac{d\sigma(E,T)}{dT} \nu(T) dT,$$

where E is kinetic energy of the projectile, T is kinetic energy transferred to the recoil atom, $T_d$ is the displacement energy, and $T_{max}$ is the highest recoil energy according to kinematics. In a modified Kinchin-Pease model, $\nu(T)$ is zero at $T<T_d$, unity at $T_d<T<2.5T_d$, and $k(T)E_d/2T_d$ at $2.5T_d<T$, where $E_d$ is "damage" energy available to generate atomic displacements by elastic collisions. $T_d$ is an irregular function of atomic number (~40 eV). The displacement efficiency, $k(T)$ drops from 1.4 to 0.3 once the PKA energy is increased from 0.1 to 100 keV, and exhibits a weak dependence on target material and temperature. The Rutherford cross-sections with Mott corrections and nuclear form-factors are used for electromagnetic elastic (Coulomb) scattering. The displacement cross-sections are calculated in the code using an improved, compared to [13], algorithm and shown in Figs. 5 and 6.

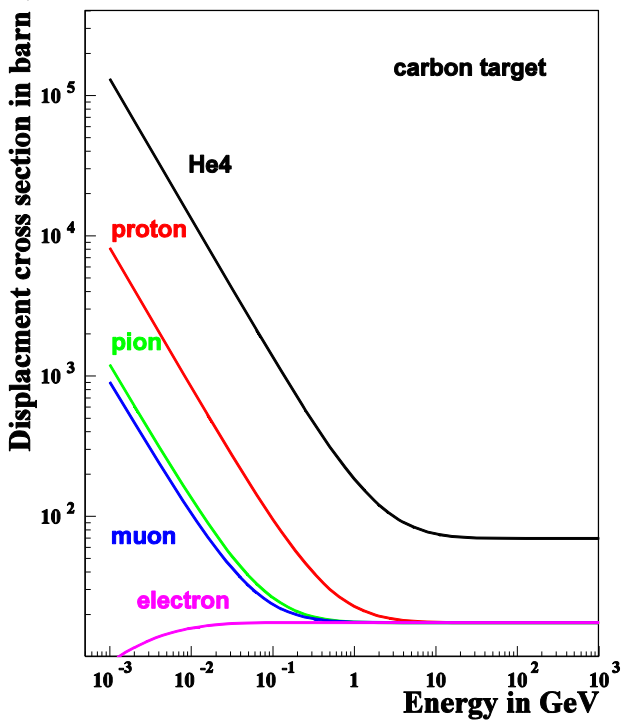

Figure 5: Displacement cross-section in carbon for various projectiles.

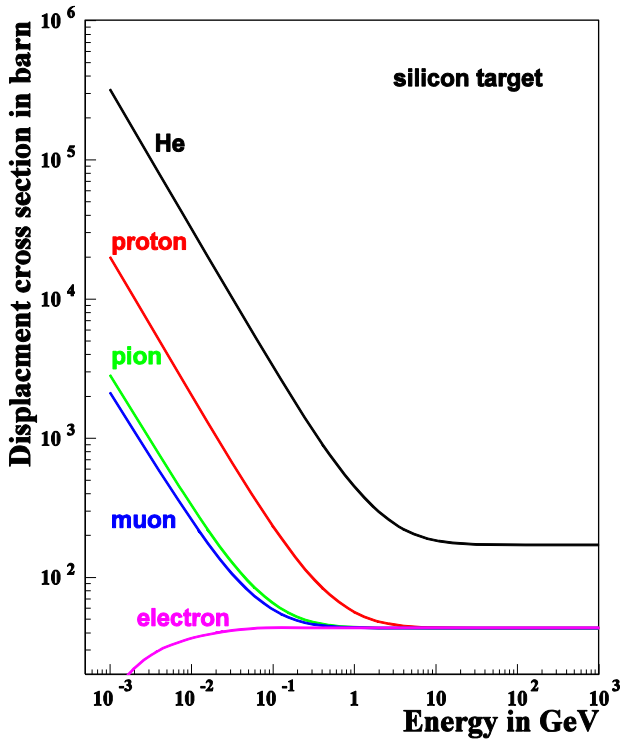

Figure 6: Displacement cross-section in silicon for various projectiles.

A comparison of the MARS15 DPA model to other DPA-capable codes showed a reasonable agreement with SRIM in a dE/dx-dominated case, a factor of three differences with DPA calculated with MCNPX and PHITS for a nuclear-dominated case and a larger disagreement for the cases where the other codes have incomplete description of the electromagnetic component of DPA [13]. Obviously, more work is needed in this area.

## BLIP BEAM TEST FOR NEUTRINO PRODUCTION TARGETS

The majority of data on radiation damage is available for reactor neutrons. Studies with hundred MeV protons have revealed that a threshold of about 0.2 DPA exists for carbon composites and graphite [14]. MARS15 studies helped realize that the BLIP beam tests at BNL with 0.165-GeV protons can emulate the neutrino production target situation for a 120-GeV proton beam at Fermilab [13]. It turns out that despite a substantial difference in the beam energies in these cases, nuclear interactions and Coulomb scattering contribute about the same way (45-50% each) to the peak DPA in graphite targets irradiated at these two facilities. Fig. 7 shows 2D distribution of DPA produced in the test module with low-Z samples after 9 weeks of irradiation with a 165-MeV proton beam of 94 mA current with $\sigma_x = 8.92$ mm and $\sigma_y = 6.79$ mm. One can see that the 0.2 DPA level can be achieved after such irradiation. The calculated hydrogen gas production rate is about $2 \times 10^{12}$ cm$^{-3}$ s$^{-1}$.

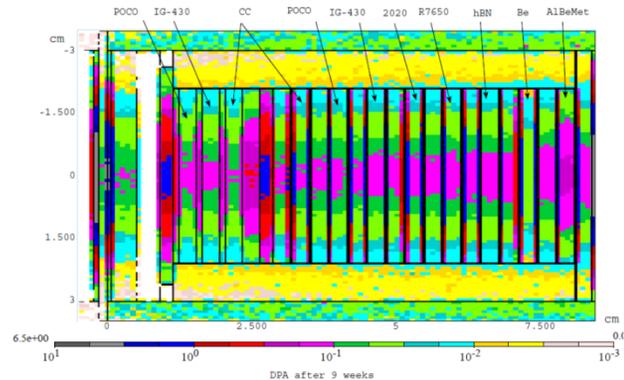

Figure 7: Calculated DPA isocontours in the BLIP module with eleven low-Z samples.

## SUMMARY

Recent developments in the MARS15 physics models, such as nuclide production, decay and transmutation and all-component DPA modelling for arbitrary projectiles in the 1 keV to 10 TeV energy range, add new capabilities to the code crucial in numerous applications with high-intensity high-power beams. Some discrepancies in DPA rate predictions by several codes, relation of DPA and H/He production rates to changes in material properties, as well as corresponding experimental studies at energies above a hundred of MeV are the areas requiring further efforts.

## ACKNOWLEDGMENTS

I would like to thank my colleagues P. Aarnio, J.D. Cossairt, K. Gudima, I. Rakhno and S. Striganov for valuable contributions to the developments described in this paper.